%% file: LESP_v2.tex
\documentclass{jfm_arxiv}
\usepackage{graphicx}
\usepackage{newtxtext}
\usepackage{newtxmath}
\usepackage{natbib}
\usepackage[hidelinks]{hyperref}
\usepackage{soul}
\usepackage[dvipsnames]{xcolor}
\usepackage[nameinlink,noabbrev]{cleveref}
\newcommand{\RomanNumeralCaps}[1]
\linenumbers

\newcommand{\kindex}[2]{\ensuremath{{#1}_{\scalebox{0.5}{#2}}}}
\newcommand{\U}{\textrm{U}}
\newcommand{\Uinf}{\kindex{\U}{$\infty$}}
\newcommand{\rle}{\kindex{r}{le}}

\newcommand{\xt}{\xi}
\newcommand{\yt}{\eta}
\newcommand{\zt}{\zeta}

\renewcommand{\upi}{\mathrm{i}}
\newcommand{\upe}{\mathrm{e}}

\def\d#1_#2{\frac{\displaystyle \mathrm{d}#1}{\displaystyle\mathrm{d}#2}}

\usepackage[locale=UK, number-mode=math, unit-mode=text, per-mode=symbol, number-unit-product=\ ]{siunitx}[=v2]

\usepackage{commath}
\hypersetup{
    colorlinks = false,
    urlcolor   = blue,
    citecolor  = black,
}

\graphicspath{{./figurematter/latex/figs/}{./figurematter/plots/}}

\shorttitle{Experimental quantification of unsteady leading-edge flow separation}
\shortauthor{J. Deparday, X. He, J. D. Eldredge, K. Mulleners, D. R. Williams}

\title{Experimental quantification of unsteady leading-edge flow separation}

\author{Julien Deparday\aff{1,2}\corresp{\email{julien.deparday@ost.ch}},
Xiaowei He\aff{3,4},
Jeff D. Eldredge\aff{5},
Karen Mulleners\aff{1},
\and David R. Williams\aff{3}
}

\affiliation{\aff{1}UNFoLD, Institute of Mechanical Engineering, \'Ecole Polytechnique F\'ed\'erale de Lausanne (EPFL), CH-1015 Lausanne, Switzerland
\aff{2}Present affiliation: Institute for Energy Technology, Eastern Switzerland University of Applied Science (OST), CH-8645 Rapperswil, Switzerland
\aff{3}MMAE Department, Illinois Institute of Technology, Chicago, IL 60616, USA
\aff{4}Present affiliation: School of Engineering, Brown University, Providence, RI 02912, USA
\aff{5}MAE Department, University of California  Los Angeles, Los Angeles, CA 90095, USA}

\begin{document}

\maketitle

\begin{abstract}
We propose here a method to experimentally quantify unsteady leading-edge flow separation on aerofoils with finite thickness.
The methodology relies on the computation of a leading-edge suction parameter based on measured values of the partial circulation around the leading-edge and the stagnation point location.
We validate the computation of the leading-edge suction parameter for both numerical and experimental data under steady and unsteady flow conditions.
The leading-order approximation of the definition of the leading-edge suction parameter is proven to be sufficiently accurate for the application to thin aerofoils such as the NACA0009 without a-priori knowledge of the stagnation point location.
The higher-order terms including the stagnation point location are required to reliably compute the leading-edge suction parameter on thicker aerofoils such as the NACA0015.
The computation of the leading-edge suction parameter from inviscid flow theory does not assume the Kutta condition to be valid at the trailing edge which allows us to compute its value for separated flows.
The relation between the leading-edge suction parameter and the evolution of the shear layer height is studied in two different unsteady flow conditions, a fixed aerofoil in a fluctuating free-stream velocity and a pitching aerofoil in a steady free-stream.
We demonstrate here that the instantaneous value of the leading-edge suction parameter based on the partial circulation around the leading edge is unambiguously defined for a given flow field and can serve as a directly quantitative measure of the degree of unsteady flow separation at the leading edge.
\end{abstract}

\maketitle

\section{Introduction}

The creation of the aerodynamic force of an aerofoil can be explained by the curvature taken by the flow due to the presence of the aerofoil \citep{babinsky_how_2003}.
The highest curvature of the aerofoil and the associated highest suction is typically located at the leading edge on the aerofoil's suction side.
Downstream of the leading edge, the pressure relaxes at the trailing edge as described by the Kutta condition to match the pressure from the other side of the aerofoil.
If the angle of attack of the aerofoil is increased, the leading-edge streamline curvature and suction increases until a critical amount of suction that can be sustained is reached and the flow will separate.
\citet{evans_analysis_1959} demonstrated a correlation between the maximum suction value of the leading-edge surface pressure coefficient and the steepness of the downstream adverse pressure gradient at stall.
This correlation is used as a stall onset criterion for incompressible flows in the dynamic stall model developed by \citet{leishman_semi-empirical_1989}.
This stall onset criterion has been adapted and improved for low Mach number flows ($Ma < 0.3$) \citep{sheng_new_2006,sheng2008modified}.
The idea of a maximal amount of supported leading-edge suction at stall is also employed as a vorticity release criterion for discrete vortex methods \citep{ramesh_discrete-vortex_2014,hou_machine-learning-based_2019,narsipur_variation_2020}.
\citet{saini2021leading} used this leading-edge suction criterion to detect vortex shedding, together with a leading-edge flow sensing method to control aerofoils in unsteady flow.
The leading-edge suction parameter is also a reliable tool to forecast the load response of an incident gusts and other external disturbances \citep{Darakananda_data_2018,LeProvost_2021}.

In the past decade, significant effort has been invested to derive a unique critical value based on the leading-edge suction, which is independent of the aerofoil shape and kinematics, that would allow for a generalisable leading-edge stall onset criterion \citep{xia_unsteady_2017,narsipur_low-order_2018,ramesh_leading-edge_2018,deparday_modeling_2019,darakananda_versatile_2019}.
To be universally used, this criterion would ideally be mathematically defined but should also be easily applicable in simulations and experiments.
A common way to define the critical leading-edge suction criterion introduced by \citet{ramesh_discrete-vortex_2014} is based on thin-aerofoil theory, where suction at the leading edge is related to the first term of the Fourier series \kindex{A}{0} defining the camberline~\citep{katz_low-speed_2004}.
\citet{ramesh_leading-edge_2018} numerically evaluated a critical value of the first Fourier coefficient when skin friction reaches a zero value at the leading edge, which happens when the flow starts to separate and a leading-edge vortex forms.
\citet{deparday_modeling_2019} showed experimentally, by integration of pressure at the leading edge, that the first Fourier coefficient reaches a maximum just before the shear-layer starts to roll up to create a leading-edge vortex.
These two studies, confirmed by \citet{Kay_LESP2021}, also showed that the observed critical maximum value of the leading-edge suction parameter depends on the unsteadiness of the aerofoil's motion.
Flow separation is a gradual process in time and space.
Under highly unsteady flow conditions, the separation point undergoes large excursions and its location does not necessarily coincide with the location of zero-skin friction \citep{haller2004exact,klose2020kinematics}.
Flow separation should not be regarded as a binary state that switches when the leading edge suction exceeds a specific threshold value, but rather as a continuous process.

The estimation of the leading-edge suction parameter based on thin aerofoil theory is limited to thin aerofoils with small camber.
Furthermore, in thin aerofoil theory, the leading-edge curvature is infinite, yielding an infinite flow velocity, hence infinite pressure and circulation at the leading edge, and the Kutta condition is imposed at the trailing edge.
Any influence of gradual progression of trailing edge separation on the leading-edge suction is thus not taken into account in the classic thin-aerofoil theory.
A first-order modification based on a spatial averaged value of the shear layer height substantially improves the prediction of the leading-edge suction for an unsteadily pitching aerofoil~\citep{deparday_modeling_2019}.

The flow around the leading edge can be modelled as a potential flow around a parabola, to avoid the pressure and velocity discontinuities encoutered using classical thin aerofoil theory.
The potential flow around a parabolic leading edge can be assimilated as the flow past a flat plate using conformal mapping \citep{van_dyke_second-order_1956}.
This method can be used to estimate the incoming flow velocity and the angle of attack based on the position of the stagnation point \citep{saini2018leading}.
The inner region close to the parabolic leading edge can be connected to an outer region using asymptotic matching \citep{van_dyke_second-order_1956}.
This asymptotic matching method allows for the evolution of the boundary layer at the leading edge to be included \citep{degani_unsteady_1996,morris_stall_2013}.
The strategy was recently adopted by \citet{ramesh_leading-edge_2020} to define the leading-edge suction based on the local flow field at the leading-edge and linked to the first term of the Fourier coefficient of the thin aerofoil theory.
\Citet{saini2021leading} also used it as a leading-edge flow sensing method, estimating the angle of attack and incoming flow speed.
Yet, this approach still relies on a Kutta condition at the trailing edge.

Alternatively, we can model the flow passing the leading edge of an aerofoil as an inviscid flow passing an edge.
Mathematically, it is possible to calculate a unique value for the circulation created by the flow passing around an edge, which is called the partial circulation by \citet{inviscidbook}.
This partial circulation can be measured experimentally as demonstrated by \citet{he_surging_2020}.
Compared to the asymptotic matching method, the partial circulation based approach only focuses on the leading-edge region and does not require information about the total circulation around the entire aerofoil, nor does it require the Kutta condition to be met.
This motivates us to further explore the potential of the partial circulation to identify unsteady leading-edge flow separation on a variety of aerofoil profiles.

In this paper, we will first present the mathematical definitions of the leading-edge partial circulation $\kindex{\varGamma}{p}$, and the leading-edge suction parameter $\sigma$, for the flow around the leading edge of an aerofoil.
Based on direct numerical simulations of the flow around a thin aerofoil (NACA0009) for a steady configuration \citep{asztalos_stall_2020}, we will demonstrate that the leading-edge suction parameter is uniquely defined and identify to what extent the obtained values are independent of the size of the leading-edge integration region.
Then, we will demonstrate the robustness of the leading-edge suction parameter definition by computing it directly from experimental data.
Two example data sets with flow separation, are considered including flow field measurements around a fixed thin (NACA0009) aerofoil in a fluctuating (surging) free-stream and a pitching thick (NACA0015) aerofoil in a steady free-stream.
By using the surging free-stream data set, we will demonstrate that the leading-edge suction parameter can be predicted based on potential flow theory even when the flow is detached, by considering the shear layer height in the leading-edge region.
Finally, we will use the observed direct relationship between the leading-edge suction parameter value and the height of the shear layer to quantify the flow separation for a pitching aerofoil undergoing dynamic stall.

\section{Definition of partial circulation and leading-edge suction parameter}

\subsection{From parabola flow to edge flow}

We approximate the leading edge of the aerofoil by a parabola with a radius of curvature \rle, described mathematically by
$\yt = \xt^{2}/(2\rle)$.
We use the complex form of coordinates $\zt = \xt + \upi \yt$.
The $\yt$ axis represents the line of symmetry of the parabola, and $\zt = 0$ is always the nose of the parabola (\cref{fig:contour}).

The complex conjugate velocity in the $\zt$ plane is derived from the complex potential $F$ (see \citet{saini2018leading,ramesh_leading-edge_2020}; details of the derivation using the current notation are provided in \cref{app:Parabola}) and is as follows
\begin{equation}
w(\zt) = -\upi U \left[ 1 - \frac{1 + \upi a \sqrt{2/\rle}}{\left( 1 + \upi 2 \zt/\rle\right)^{1/2}}\right],
\label{eq:velo}
\end{equation}
with $a$, the position of the stagnation point on the parabola.
This flow is uniform at a velocity $U$ at leading order at large distances $|\zt| \gg \rle$ from the nose of the parabola.
The inverse square root in the second term in \cref{eq:velo} shows that the disturbance to this uniform flow is equivalent to the flow about a sharp edge at $\zt = \upi \rle/2$ \citep{inviscidbook}.
Although this effective edge is interior to the parabola, this fact becomes increasingly irrelevant at large distances from the nose: the influence of nose curvature becomes negligible at such distances.

Let us compare this second term with the generic form of complex velocity for an edge flow \citep{inviscidbook},
\begin{equation}
\frac{\upi \sigma \kindex{n}{0}^{-1/2}c^{1/2}}{\left(\zt-\kindex{\zt}{0}\right)^{1/2}},
\end{equation}
where $\sigma$ is the leading-edge suction parameter, $c$ is the chord length of the aerofoil, $\kindex{\zt}{0}$ is the edge coordinate, and $\kindex{n}{0}$ is the unit vector directed outward from the edge (and tangent to the plate).
By comparing these, we can conceive an \emph{effective edge}, with an associated strength and orientation.
This comparison leads to
\begin{equation}
\sigma = \frac{Ua}{c^{1/2}} \left( 1 + \frac{1}{2} \frac{\rle}{a^{2}}\right)^{1/2}
\label{eq:lesp}
\end{equation}
and
\begin{equation}
\kindex{n}{0} = -\upi \upe^{\upi 2\phi}, \quad \phi = \arctan \left(\sqrt{\rle/2}/a\right).
\end{equation}
\Cref{eq:lesp} gives us a mathematically well-defined leading-edge suction parameter for a flow around a leading edge, using a potential flow around a parabola that is equivalent to an edge flow.
To compute this leading-edge suction parameter from velocity field data in an unambiguous way, we will introduce the concept of partial circulation.

\subsection{Partial circulation}

\begin{figure}
\centering
\includegraphics[width=0.4\textwidth]{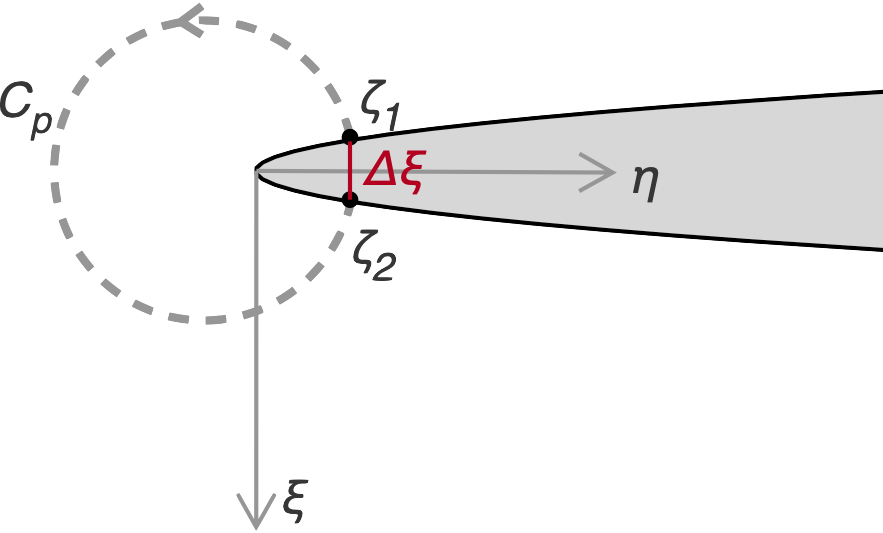}
\caption{Contour of integration for partial circulation, $\kindex{\varGamma}{p}$.}
\label{fig:contour}
\end{figure}

Suppose we compute the velocity along a contour $\kindex{C}{p}$ joining two points, $\kindex{\zt}{1}$ and $\kindex{\zt}{2}$, situated on the parabola, as seen in \cref{fig:contour}.
We will call this the partial circulation, $\kindex{\varGamma}{p}$, as in \citet{inviscidbook}, defined in complex form as
\begin{equation}
\kindex{\varGamma}{p} = \int_{\kindex{C}{p}} w\,\mathrm{d}\zeta;
\end{equation}
the imaginary part of the integral, which represents the volume flow rate across the contour, vanishes due to conservation of mass as long as the parabola is stationary.
This integral can be easily evaluated by replacing $w$ with $\mathrm{d}F/\mathrm{d}z$, so that
\begin{equation}
\kindex{\varGamma}{p} = F(\zeta_{2}) - F(\zeta_{1}),
\end{equation}
which represents the difference in velocity potentials since the imaginary part is zero.
It is the same along any two contours joining these two points.
This holds because the flow is irrotational.
(If there had been vorticity in this parabola, then pairs of contours that enclose such vorticity would obtain different results, different precisely by the circulation of the enclosed vorticity.)
Let us further assume that $\kindex{\zt}{2}$ and $\kindex{\zt}{1}$ are situated symmetrically about the axis of symmetry of the parabola so that their $\xt$ coordinates are equal and opposite.
We will denote these coordinates by $\Delta\xt/2$ and $-\Delta\xt/2$.
Thus, by conformal mapping, we get 
\begin{equation}
\kindex{\varGamma}{p} = -\Delta\xt U a \sqrt{\frac{2}{\rle}}.
\label{eq:gampart}
\end{equation}
Comparing \cref{eq:gampart} with \cref{eq:lesp}, we can relate the leading-edge suction parameter to the partial circulation:
\begin{equation}
\sigma = - \frac{\kindex{\varGamma}{p}}{\Delta\xt} \sqrt{\frac{\rle}{2 c}}\left( 1 + \frac{1}{2} \frac{\rle}{a^{2}}\right)^{1/2}.
\label{eq:lesp1}
\end{equation}

This expression still contains the stagnation point parameter $a$.
If the position of the stagnation point is known, we can determine the leading-edge suction parameter using \cref{eq:lesp1}.
If $\rle \ll a^{2}$, we can simplify \cref{eq:lesp1} and the leading-edge suction parameter is determined by
\begin{equation}
\sigma = - \frac{\kindex{\varGamma}{p}}{\Delta\xt} \sqrt{\frac{\rle}{2 c}} + O(\rle/a^{2}).
\label{eq:lesp2}
\end{equation}

In summary, we can estimate the leading-edge suction parameter to a leading order from \cref{eq:lesp2} by simply computing the partial circulation on any contour and dividing by the lateral distance between the two symmetric endpoints of the contour.
If the location of the stagnation point is known, we can improve this estimate using \cref{eq:lesp1}.

We will use and test these two formulas, \cref{eq:lesp2} and \cref{eq:lesp1}, using simulations and experimental data, respectively on a static thin aerofoil (NACA0009) and on a pitching thicker aerofoil (NACA0015).

\section{Validation of the leading-edge suction parameter}

\subsection{Method to compute the partial circulation at the leading edge}
\label{subsec:defCirc}

To estimate the leading-edge suction parameter $\sigma$, in \cref{eq:lesp1,eq:lesp2}, an estimate of the partial circulation $\kindex{\varGamma}{p}$, is required.
In this paper, we will compute the partial circulation by integrating the vorticity in experimentally obtained flow field snapshots along a circular contour enclosing the leading edge (\cref{fig:contour}).
The contour does not have to be circular and can in principle have various shapes.
To facilitate the computation and to reduce the influence of measurement noise and the influence of sparse or missing measurement data within the boundary layer, we selected a circular contour that intersects perpendicularly with the aerofoil surface.
Any flow perpendicular to the contour would not contribute to an increase in the partial circulation.
If the flow is attached in the leading edge region, the flow in the boundary layer is parallel to the surface, hence normal to the integration contour.
The result of the line integral of the tangential velocity is thus only marginally affected by missing measurement data in the boundary layer due to surface reflections or shadow regions.
The assumption that the flow is normal to the integration contour close to the aerofoil surface remains true for trailing-edge separation, but fails when the flow separates from the leading edge.
The influence of trailing edge separation on the values of the partial circulation change will be quantified later.
When leading edge flow separation occurs, vorticity is ejected away from the aerofoil surface in the regions where measurement data is present.
The region near the surface where measurement data is missing is small and the majority of the vorticity during leading-edge separation is concentrated in the shear layer that is well resolved for the cases presented in this paper.

If the circular integration contour has to intersect perpendicularly with the aerofoil contour, it suffices to fix the chord-wise position of the intersections or the endpoints to deduce the radius of the circular integration contour and its centre.
The sole parameter that characterises the integration contour is thus the chord-wise position of the endpoints, which should not have an influence on the estimation of the leading-edge suction parameter.
We will demonstrate next that this holds true for chord-wise locations within the first \SI{10}{\percent} of the aerofoil.

\subsection{Validation of the leading-edge suction parameter on a static thin aerofoil}
\label{subsec:validationLESP}

\begin{figure}
\centering
\includegraphics{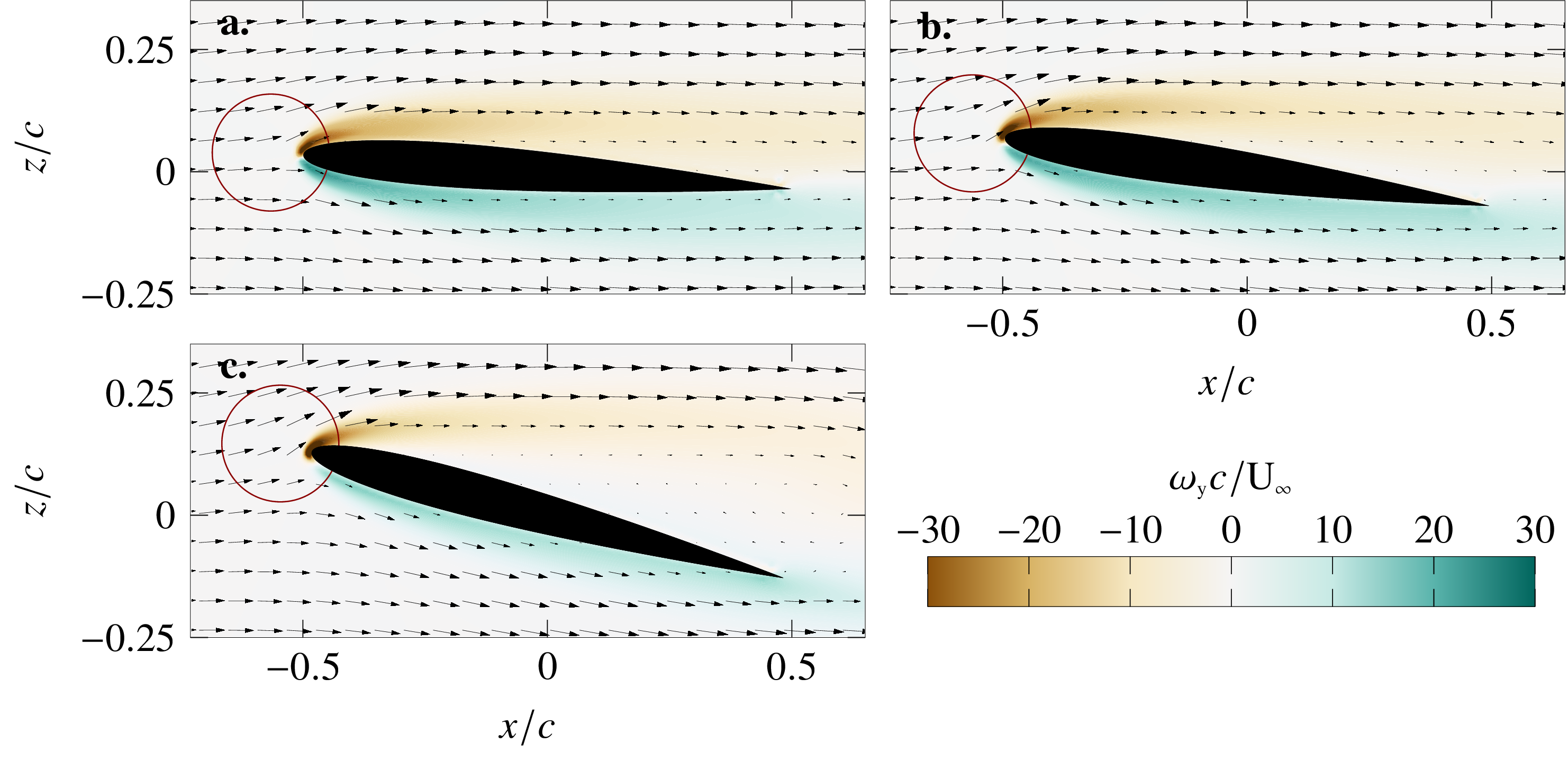}
\caption{Averaged vorticity and velocity field from DNS results on a NACA0009 at an angle of attack of $\alpha = \ang{4}$ (a.), $\alpha = \ang{8}$ (b.), and $\alpha = \ang{15}$ (c.).
The circular arc represents the contour used for the partial circulation and encloses \SI{5}{\percent} of the chord.
}
\label{fig:PartialCirc_flow}
\end{figure}

To verify the independence of the leading-edge suction parameter on the size of the integration contour, we first compute the partial circulation and suction parameter here for a simple case of a thin static aerofoil with laminar flow.
The partial circulation is calculated on the results of a direct Navier-Stokes simulation (DNS).
In this first example, we use the DNS results that do not suffer from missing data or experimental noise, which would alter the results and hinder the conclusions on the validation of the leading-edge suction parameter.
\Cref{fig:PartialCirc_flow} shows the velocity and vorticity fields for a NACA0009 aerofoil at angles of attack $\alpha = \ang{4}$ (\cref{fig:PartialCirc_flow}a), $\alpha = \ang{8}$ (\cref{fig:PartialCirc_flow}b), and $\alpha = \ang{15}$ (\cref{fig:PartialCirc_flow}c).
The displayed circular arc contour used to compute the partial circulation encloses \SI{5}{\percent} of the chord.
The results of the flow field are obtained from solving the two-dimensional Navier-Stokes equations with an immersed boundary projection method at a Reynolds number $Re=500$.
The original computational domain includes the flow field from $0.5c$ upstream of the leading edge to $5c$ downstream of the trailing edge.
Depending on the angle of attack and the free-stream conditions, the aerofoil can experience attached flow, partial separation, static stall, etc.
The details of the DNS process are documented in \citet{asztalos_stall_2020}.
Here, we use the DNS results to test the robustness of the leading-edge suction parameter $\sigma$ when the flow is fully attached at $\alpha = \ang{4}$, when the flow starts to separate near the trailing-edge at $\alpha = \ang{8}$, and when the flow is fully separated at $\alpha = \ang{15}$.

\begin{figure}
\centering
\includegraphics{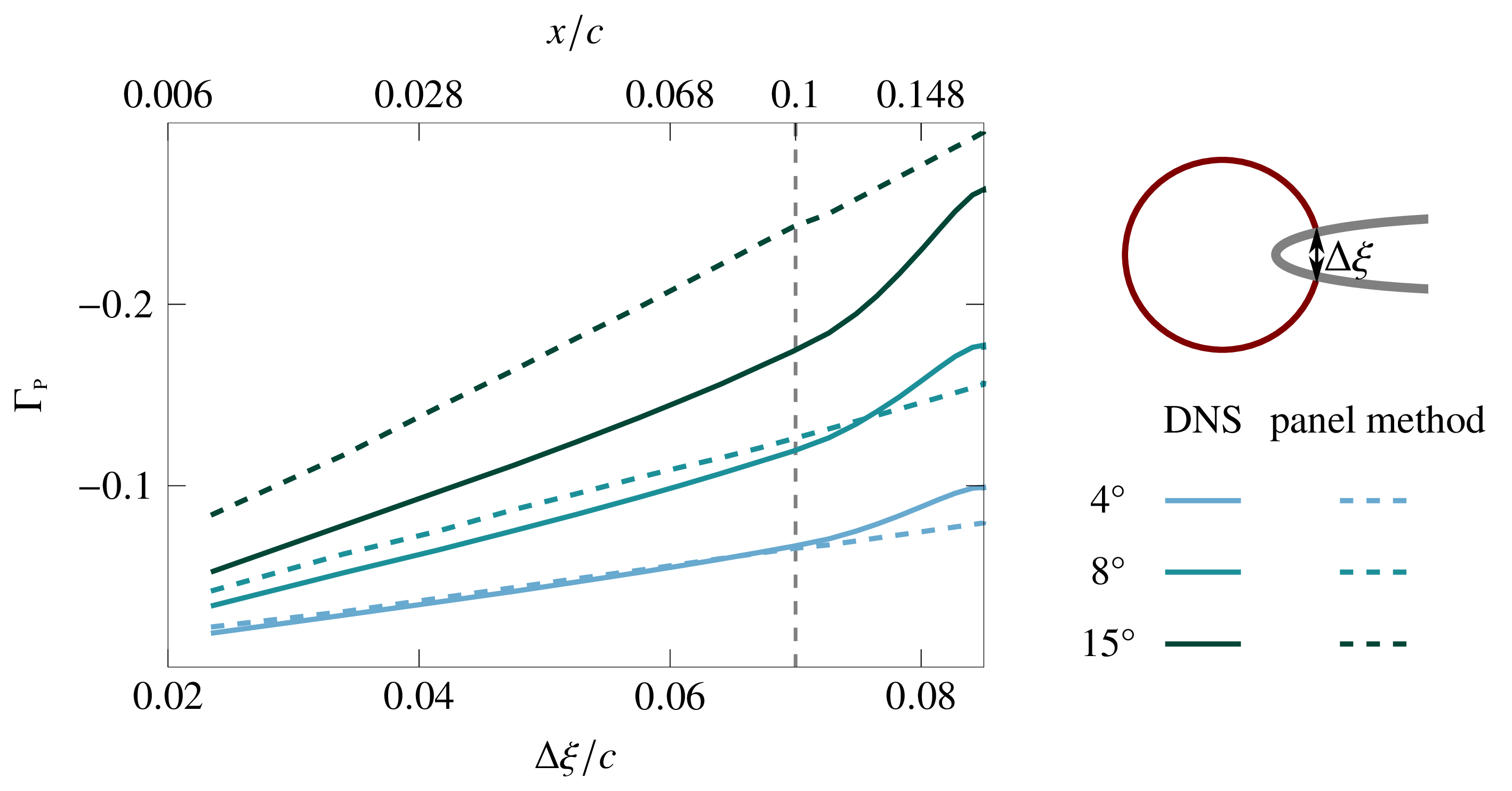}
\caption{Partial circulation $\kindex{\varGamma}{p}$ as a function of the distance between the endpoints of the integration contour for a NACA0009 at various angles of attack: $\alpha=$\ang{4}, \ang{8}, \ang{15}.
The solid lines indicate data obtained using DNS.
The dashed lines indicate values obtained using a panel method.
The axis on top indicates the chord-wise location of the endpoints of the integration contours.}
\label{fig:PartialCirc}
\end{figure}

For a thin aerofoil such as the NACA0009, $\rle \ll a^{2}$ and the leading-edge suction parameter $\sigma$ can be calculated using the simplified \cref{eq:lesp2}.
For a fixed angle of attack and steady flow conditions, the stagnation point does not change and the only terms that vary in \cref{eq:lesp2} when we vary the integration contours are the partial circulation, \kindex{\varGamma}{P}, and the thickness of the aerofoil at the endpoints, $\Delta\xt$.
The variation of the partial circulation with increasing size of the integration contours is presented as a function of the thickness of the aerofoil at the endpoints in \cref{fig:PartialCirc} for the NACA0009 at three fixed angles of attack: $\alpha=$\ang{4}, \ang{8}, and \ang{15}.
The solid lines indicate the results for the partial circulation obtained from the DNS flow field data.
The dashed lines indicate the results for the partial circulation calculated by a simple panel method without boundary layer model and the Kutta condition at the trailing edge.
The x-axis at the bottom indicates the thickness of the aerofoil at the endpoints and the x-axis on the top indicates the portion of the chord that is enclosed by the integration contour.
For all angles of attack, the magnitude of the partial circulation increases when the contour increases and encloses a larger portion of the leading edge.
The negative sign of the partial circulation demonstrates that the circulation at the leading edge is associated with suction.
As the suction and aerodynamic force increase with increasing angle of attack, the magnitude of the partial circulation increases with the angle of attack.

The partial circulations for the three angles of attack increase linearly with the thickness of the aerofoil until the contour encloses approximately \SI{10}{\percent} of the chord.
A constant slope of the \kindex{\varGamma}{P} versus $\Delta\xt$ curve indicates a constant value of the leading-edge suction parameter.
Based on the results presented in \cref{fig:PartialCirc}, we can conclude that the leading-edge suction parameter based on the DNS flow field data would yield a unique value for a fixed angle of attack independent on the size of the integration contour, as long as the contour encloses no more than the first \SI{10}{\percent} of the chord.

The partial circulations calculated by the panel method are presented by the dashed lines in \cref{fig:PartialCirc}.
At an angle of attack of $\alpha = \ang{4}$, when the flow is attached, the partial circulation from the panel method fits well the DNS value until \SI{10}{\percent} of the chord.
For contours enclosing more than \SI{10}{\percent} of the chord, the partial circulation obtained with the panel method continues to increase linearly with $\Delta\xt$, but the DNS based partial circulation shows a steeper increase.
At an angle of attack of $\alpha = \ang{8}$, the flow is separated in the trailing edge region but not at the leading edge.
The difference between both methods is larger than at $\alpha = \ang{4}$ but stays constant for the first \SI{10}{\percent} of the chord.
In this region the slopes of the curves and thus the suction parameter for both methods are in close agreement.
At an angle of attack of $\alpha = \ang{15}$, the flow is separated at the leading edge.
The magnitude of the partial circulation and the slope of the \kindex{\varGamma}{P} versus $\Delta\xt$ curve are higher for the panel method than for the DNS results, as the flow is considered attached in the panel method.

In conclusion, if the partial circulation is calculated using a contour enclosing the first \SI{10}{\percent} of the chord, the leading-edge suction parameter yields a unique value which can be measured and estimated without information about the position of the stagnation point.
When flow separation occurs, the leading-edge suction parameter estimation from the panel method deviates from the directly computed values and motivates us to examine more closely the influence of the shear-layer in the next section.

\section{Influence of shear layer on the leading-edge suction parameter}

\subsection{General idea}
\label{subsec:LESPshearlayer}
\begin{figure}
	\centering
	\def\svgwidth{0.5\linewidth}
	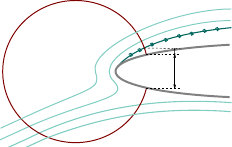
	\caption{Sketch of the leading-edge region with the presence of the shear layer (dark green). A contour for the partial circulation is represented in red. The presence of the shear layer at the extremity of the contour increases virtually the thickness of the aerofoil by $\delta \kindex{\xt}{SL}$.}
	\label{fig:SketchSL}
\end{figure}

The shear layer is a region of high vorticity at the interface between low velocity, separated flow and the high velocity free-stream flow.
In a time-averaged sense, the shear layer acts as a separation line across which there is no fluid transport and it can be seen as a virtual aerofoil edge, with the recirculation region as part of this virtual aerofoil.
As the shear layer would increase the thickness of the virtual aerofoil, the denominator $\Delta\xt$ in \cref{eq:lesp1} would increase by the height of the shear layer $\delta \kindex{\xt}{SL}$ at the endpoints of the contour used for the partial circulation (\cref{fig:SketchSL}).

The presence of the shear layer would not affect the vorticity within the contour, if we assume $\delta \kindex{\xt}{SL} \ll \Delta\xt$.
The change of the intensity of the partial circulation $\kindex{\varGamma}{p}$ is assumed to be much lower than the change of thickness of the aerofoil $\Delta\xt + \delta \kindex{\xt}{SL}$.
In \cref{eq:lesp1} and \cref{eq:lesp2}, the only term changing due to the presence of the shear layer is the thickness of the aerofoil at the contour.
The partial circulation, the stagnation point position and the radius of the leading edge would stay the same as if the flow was attached and in the absence of a trailing edge shear layer (\cref{eq:lespdx}):
\begin{equation}
\sigma = - \frac{\kindex{\varGamma}{p}}{\Delta\xt + \delta \kindex{\xt}{SL}} \sqrt{\frac{\rle}{2 c}}\left( 1 + \frac{1}{2} \frac{\rle}{a^{2}}\right)^{1/2}.
\label{eq:lespdx}
\end{equation}

Similarly to \cref{eq:lesp2}, for small $\rle/a^2$, we can simplify \cref{eq:lespdx} and the corrected leading-edge suction parameter is determined by

\begin{equation}
\sigma = - \frac{\kindex{\varGamma}{p}}{\Delta\xt + \delta \kindex{\xt}{SL}} \sqrt{\frac{\rle}{2 c}} + O(\rle/a^{2}).
\label{eq:lespdx2}
\end{equation}

Next, we will verify the proposed correction for the estimation of the leading-edge suction parameter for a thin aerofoil at a fixed angle of attack that is subjected to stream-wise flow perturbations.

\subsection{Thin aerofoil with incoming flow perturbations}
\label{subsec:SLflowpert}

\begin{figure}
\centering
\includegraphics{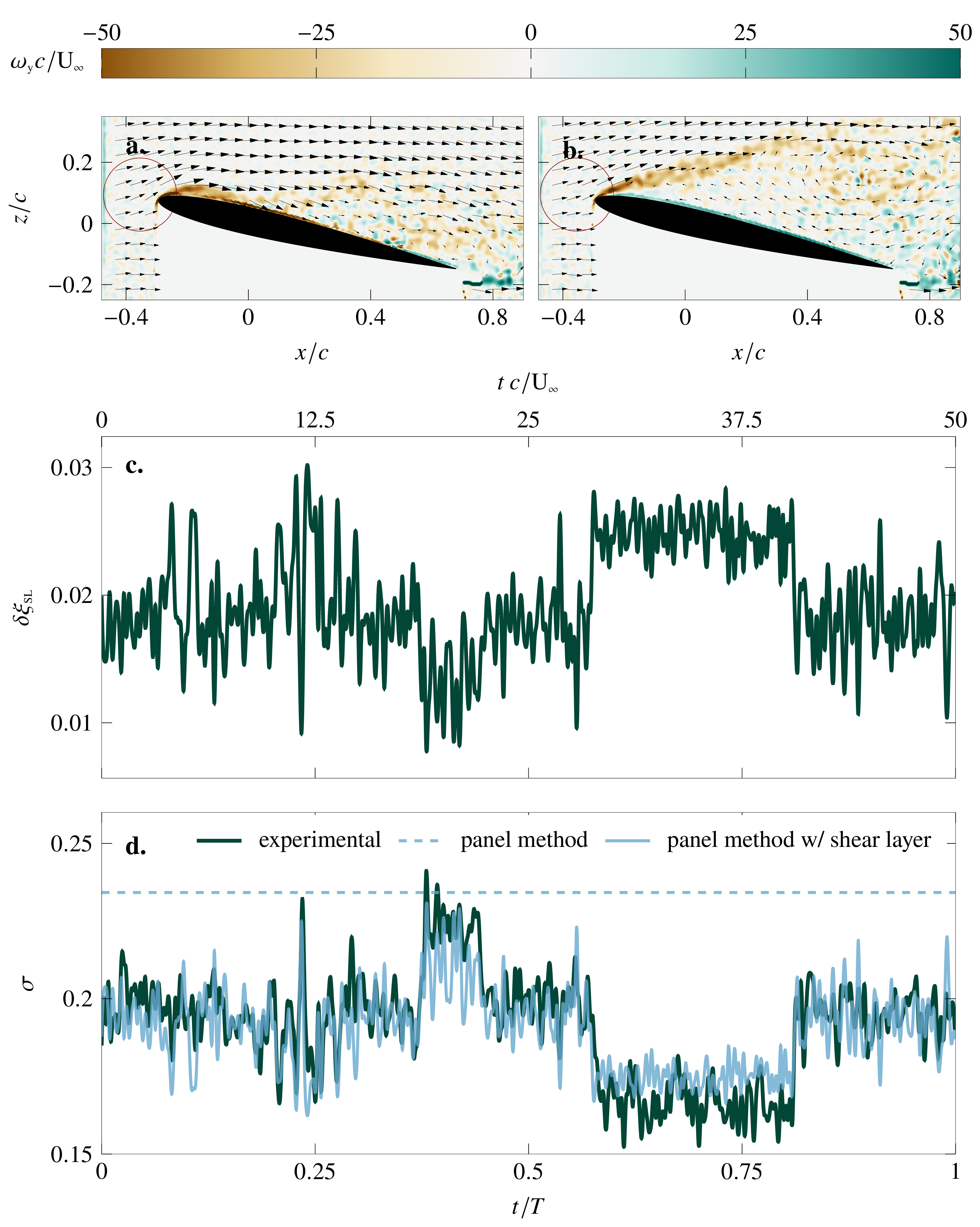}
\caption{Two snapshots presenting the flow field and vorticity field around a NACA0009 in a surging flow, at \ang{15} angle of attack, when the flow is considered attached (a.) and detached (b.). Height of the shear layer at the extremity of the contour (c.) and evolution of the leading-edge suction parameter (d.). The "experimental" curve corresponds to the leading-edge suction parameter calculated using the experimental flow field, the "panel method" curve corresponds to the constant value obtained at a fixed angle of attack and "panel method w/ shear layer" takes into account the corrected thickness of the aerofoil $\Delta\xt + \delta \kindex{\xt}{SL}(t)$.}
\label{fig:surgingFlowLE_SL}
\end{figure}


As a first validation, we consider a NACA0009 aerofoil at a fixed angle of attack $\alpha=\ang{15}$ under surging flow conditions.
The experiments were conducted in the Andrew Fejer Unsteady Wind Tunnel at the Illinois Institute of Technology.
The semi-2D aerofoil has a chord length of \SI{0.245}{\metre} and a wing span of \SI{0.6}{\metre} and was placed in the middle of the test section which has a \SI{0.61x0.61}{\metre} cross section.
With a mean flow speed of \SI{5.8}{\metre\per\second}, the chord based Reynolds number is $Re=\num{9.8e4}$.
The flow was perturbed by a louver mechanism at the downstream end of the test section, so that the flow could surge in the longitudinal direction.
The details of the experimental setup and the surging flow setup can be found in \citet{he_2020_spectral,he_surging_2020}.
Time-resolved PIV was carried out to assess the experimental unsteady leading-edge suction parameter $\sigma$.
The PIV velocity vector field was constructed on a \num{320x200} grid over a \SI{381x238}{\milli\metre} window at a sampling rate of \SI{100}{samples\per\second}.

At \ang{15} angle of attack, the flow is fully separated over the aerofoil and the shear layer is detached in the leading-edge region.
In response to the variations in the stream-wise velocity, the shear layer varies between reattached around the leading edge and fully separated as demonstrated in the snapshots in \cref{fig:surgingFlowLE_SL}a,b.
In general, acceleration of the free-stream leads to the reattachment of the shear layer to the upper surface of the aerofoil (\cref{fig:surgingFlowLE_SL}a) and deceleration of the free-stream causes the shear layer to fully separate (\cref{fig:surgingFlowLE_SL}b).
The shear layer height is measured in the velocity field snapshots at $0.1 x/c$, slightly behind the integration contour for the partial circulation to have a higher signal to noise ratio.
The shear layer height at $0.05x/c$, where the partial circulation is measured, is obtained by a linear interpolation \citep{deparday_modeling_2019}.
The time-series plot of the shear layer height $\delta\kindex{\xt}{SL}$ is presented in \cref{fig:surgingFlowLE_SL}c.
For most of the measurement window, $\delta\kindex{\xt}{SL}$ fluctuates about the mean value of \num{0.017}.
Between $t/T=0.37$ and $t/T=0.45$, there is a significant reduction in $\delta\kindex{\xt}{SL}$, which corresponds to the shear layer reattachment that is shown in \cref{fig:surgingFlowLE_SL}a.
We see $\delta\kindex{\xt}{SL}$ increasing between $t/T=0.6$ and $t/T=0.8$, when the shear layer fully detaches (see \cref{fig:surgingFlowLE_SL}b).

We now examine the relation between the shear layer and the leading-edge suction parameter by looking into the influence of the height of the shear layer $\delta\kindex{\xt}{SL}$.
The partial circulation $\kindex{\varGamma}{p}$ is measured at $0.05x/c$ using the contour indicated in \cref{fig:surgingFlowLE_SL}a,b.
The leading-edge suction parameter is obtained using \cref{eq:lesp2} with $\Delta \xt = 0.052c$.
The experimental results are presented in dark green in \cref{fig:surgingFlowLE_SL}d.
The shear layer height $\delta\kindex{\xt}{SL}$ and the experimental leading-edge suction parameter $\sigma$ are negatively correlated: $\sigma$ increases when $\delta\kindex{\xt}{SL}$ decreases.
In other words, the leading-edge suction is stronger when the flow is more attached and weaker when the flow is partly or entirely separated.

Motivated by the visual correlation between the evolution of the leading-edge suction parameter and the shear layer height at the leading edge, we now seek to include the influence of the experimentally determined shear layer height on the calculation of the theoretical leading-edge suction parameter.
We obtain a correction due to the viscous effect which is applied to the inviscid theory.
The theoretical leading-edge suction parameter uses the partial circulation computed from a panel method with the flow completely attached on the aerofoil and the Kutta condition at the trailing edge.
From \cref{eq:lesp2}, the resulting leading-edge suction parameter from this panel method is a constant value (blue dashed line in \cref{fig:surgingFlowLE_SL}d), since the partial circulation is solely a function of the aerofoil geometry and the angle of attack when assuming a fully attached flow.
The partial circulation $\kindex{\varGamma}{p}$ does not change for a fixed contour location ($0.05x/c$) in the panel method solution, but when we use $\Delta \xt+\delta\kindex{\xt}{SL}(t)$ (\cref{eq:lespdx}), the corrected leading-edge suction parameter varies due to the influence of the temporal evolution of the shear layer height (solid blue line in \cref{fig:surgingFlowLE_SL}d).
Once the time-resolved virtual thickness on the aerofoil is taken into account, which can also be considered as the virtual separation in the inviscid flow, the leading-edge suction parameter value deviates from the fully attached value and matches with the experimental value.
By applying the correction proposed in \cref{eq:lespdx}, the dynamics of the theoretical prediction match the dynamics of the time-resolved experimental measurements with a cross-correlation coefficient of \num{0.81} and no time delay.

The difference between the estimation of the leading-edge suction parameter from a simple panel method and the experimental values is mainly due to the occurrence of separation which is manifested in an increase in the shear layer height.
The close interplay between the leading-edge suction parameter and the shear layer height can also be exploited as a means to quantitatively identify the leading-edge flow separation.
The difference between the measured leading-edge suction parameter and its predicted value based on potential flow theory is explored next to identify the state of the flow past a pitching aerofoil.

\section{Leading-edge suction parameter on a thicker pitching aerofoil}

\subsection{Presentation of the test case}
\label{subsec:pitchingaerofoil}

We now consider the flow past a pitching aerofoil experiencing dynamic stall.
The experiments were conducted in a recirculating wind tunnel at the German aerospace center in G{\"o}ttingen.
The open jet test section of the tunnel has a nozzle end size of \SI{0.75x1.05}{\meter}.
The tunnel was operated at an incoming free stream velocity of $\Uinf = \SI{30}{\meter\per\second}$.
The span of the aerofoil was $\SI{1.05}{\meter}$ and the chord length $c=\SI{0.3}{\meter}$.
The Reynolds number based on the chord length is $Re = \num{5.5e5}$.

The aerofoil used in this dynamic stall case is a NACA0015.
The maximum thickness relative to the chord is larger than for the previous case with a NACA0009, and the radius of the leading edge is also larger ($0.025c$ instead of $0.009c$).
The aerofoil is pitched about its quarter chord axis such that the angle of attack varies sinusoidally around its static stall angle of $\kindex{\alpha}{0}=\kindex{\alpha}{ss}=\ang{20}$ with an amplitude of $\kindex{\alpha}{1}=\ang{8}$ and a frequency of $\kindex{f}{osc} = \SI{31.4}{\radian\per\second}$.
The reduced frequency $k=\pi\kindex{f}{osc}c/\Uinf$ for the aerofoil is \num{0.1}.

The unsteady aerofoil surface pressure distribution was measured by \num{36} pressure transducers distributed in the mid-span section of the aerofoil.
We calculated the aerodynamic force generated by the aerofoil by integrating the surface pressure distribution.
In addition, a two-dimensional PIV was used to measure the flow field at the mid-span of the aerofoil, with a sampling rate of \SI{1.5}{\kilo\hertz}.
Two high-speed cameras recorded images with a focus on different parts of the aerofoil.
A first camera covered the leading edge.
With the use of mirrors to project the laser sheet on both sides of the aerofoil, shaded areas were minimised, and it allowed us to properly measure the circulation at the leading edge using a contour as described in \cref{subsec:defCirc}.
A multi-grid algorithm was used to obtain the velocity vectors with a final window size of \SI{48x48}{px} and \SI{50}{\percent} overlap yielding a physical grid resolution of $\SI{1.8}{\mm}=\SI{0.0061}{c}$ in the leading-edge region.
A second camera had a much larger field of view covering the rest of the aerofoil's suction side, making it possible to visualise the evolution of the shear layer during the pitching motion.
The physical grid resolution for the second camera was of $\SI{5.4}{\mm}=\SI{0.018}{c}$.
More details of the experimental setup can be found in \citet{he_stall_2020}.

\begin{figure}
\centering
\includegraphics{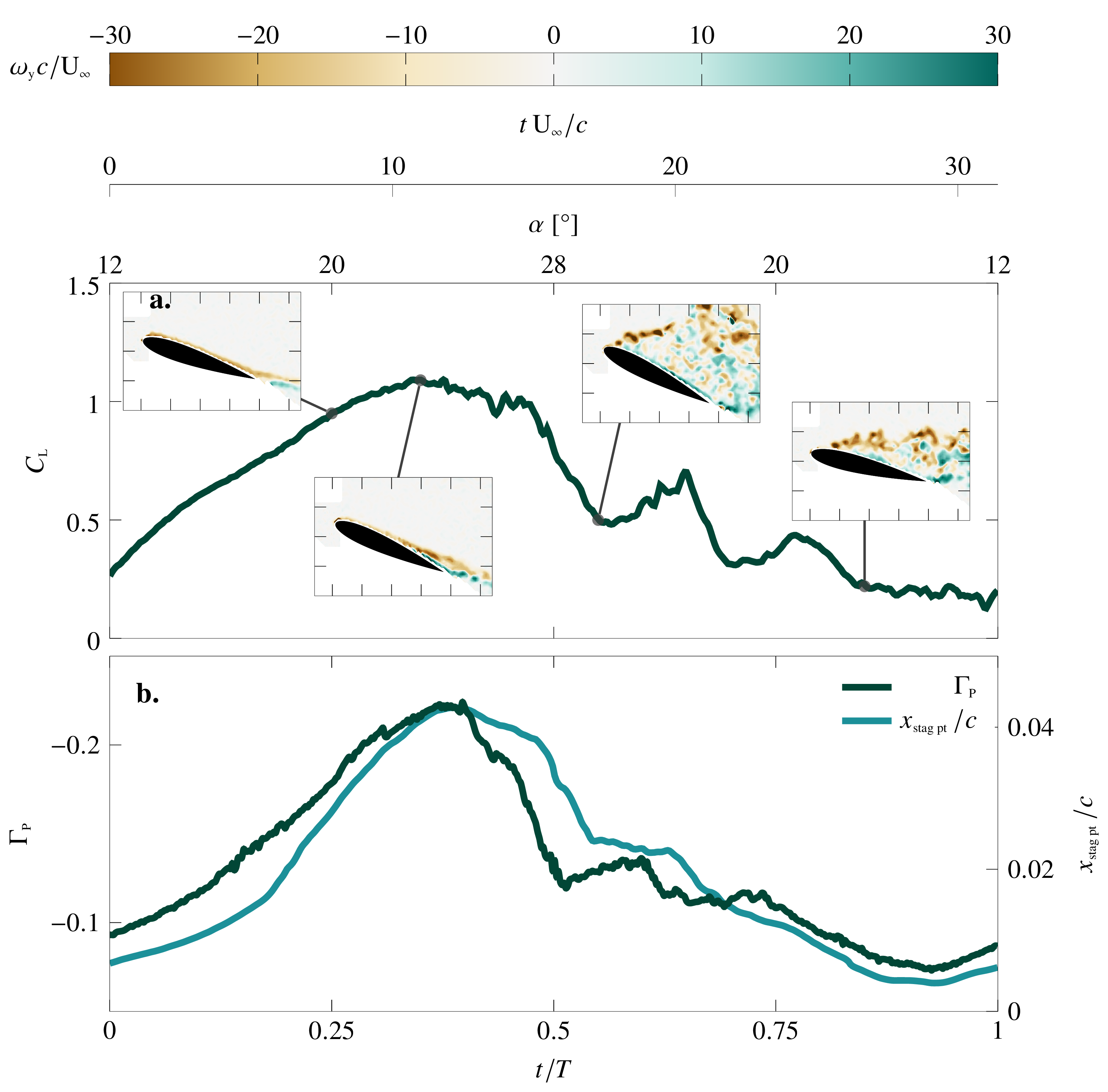}
\caption{Evolution during a pitching cycle, T, of a NACA0015 aerofoil of: a. lift evolution with four snapshots summarising the flow evolution, b. partial circulation at 0.05x/c (left axis) and position of stagnation point (right axis).}
\label{fig:pitchingOverview}
\end{figure}

\Cref{fig:pitchingOverview} presents the evolution of the lift coefficient \kindex{C}{L}, of the partial circulation at the leading edge using a contour with the endpoints at $x/c = 0.05$, $\kindex{\varGamma}{p}$, and of the chord-wise position of the stagnation point, \kindex{x}{stg}, during a sinusoidal pitching cycle, $T$.
The non-dimensional convective time starting from moment when the aerofoil is at its lowest angle of attack is indicated on the top x-axis together with the corresponding angle of attack.
In the beginning of the cycle, the lift coefficient increases approximately linearly with the angle of attack.
When the static stall angle of \kindex{\alpha}{ss}=\ang{20} is reached, the flow is still attached on the suction side of the aerofoil, as seen in the corresponding inserted snapshot in \cref{fig:pitchingOverview}a., and the lift continues to increase with increasing angle of attack.
When the lift coefficient reaches its maximum value, positive vorticity appears in the last quarter of the aerofoil, between the suction side and the negative vorticity line which highlights the shear layer.
The positive vorticity indicates the detachment of the flow, creating a recirculating region which starts at the trailing edge and grows.
The upstream limit of the recirculating region moves upstream, finally reaching the leading edge \citep{mulleners_onset_2012}.
The shear layer at the edge of this recirculating flow region rolls-up to form a primary leading-edge vortex or dynamic stall vortex.
While the recirculation region grows and the dynamic stall vortex forms, the lift coefficient stays close to its maximum value of \num{1.1} during four convective times.
Once the dynamic stall vortex separates, the lift coefficient starts decreasing.
This happens shortly before the maximum angle $\alpha=\ang{28}$ is reached for the selected pitching kinematics.
The lift coefficient quickly decreases to reach a local minimum of less than \num{0.5} when the flow on the suction side is completely detached as seen in \cref{fig:pitchingOverview}a. at $t/T=0.6$.
At the end of the pitching motion, the flow reattaches starting from the leading edge to the trailing edge.

The circulation created at the leading edge is negative.
Lower negative values at the leading edge indicate a stronger circulation and an increase in the overall lift.
The circulation generated at the leading edge (\cref{fig:pitchingOverview}b) follows the same overall evolution as the lift coefficient but the correlation coefficient between both is not one.
For example, the maximum absolute value of the circulation at the leading edge is reached after the maximum lift coefficient, when the leading-edge vortex is being formed.
While the lift coefficient stays at its maximum value for about three convective times, the partial circulation decreases immediately after reaching its maximum.
The post-stall local maxima in circulation and lift do not occur at the same time either.

The chord-wise position of the stagnation point (\cref{fig:pitchingOverview}b, right axis) has been identified as the stagnation pressure location from the spatially interpolated surface pressure distribution that was measured simultaneously with the PIV.
The chord-wise position of the stagnation point is, at the beginning of the sinusoidal motion, at less than \SI{1}{\percent} of the chord from the leading edge.
The position of the stagnation point starts close to the leading edge for low angles of attack and moves further downstream towards the trailing edge when the angle of attack increases.
The leading-edge circulation reaches a maximum value when the leading-edge stagnation point has reached its most downstream position of $x/c=0.04$.
The stagnation point position does not decrease as fast as the partial circulation just after its maximum is reached.

The location of the stagnation point oscillates around \num{0.02}c which is of the same order of magnitude as the radius of the leading edge of the aerofoil (\num{0.025}c).
The second term of \cref{eq:lesp1} includes the ratio between the stream-wise position of the stagnation point, $a^2$, and the radius of the leading edge, $\rle$.
This term is not negligible here, and should be taken into account to calculate the leading-edge suction parameter.

\subsection{Comparison of the leading-edge suction parameter with potential flow estimations}
\label{subsec:LESPpitching}

\begin{figure}
\centering
\includegraphics{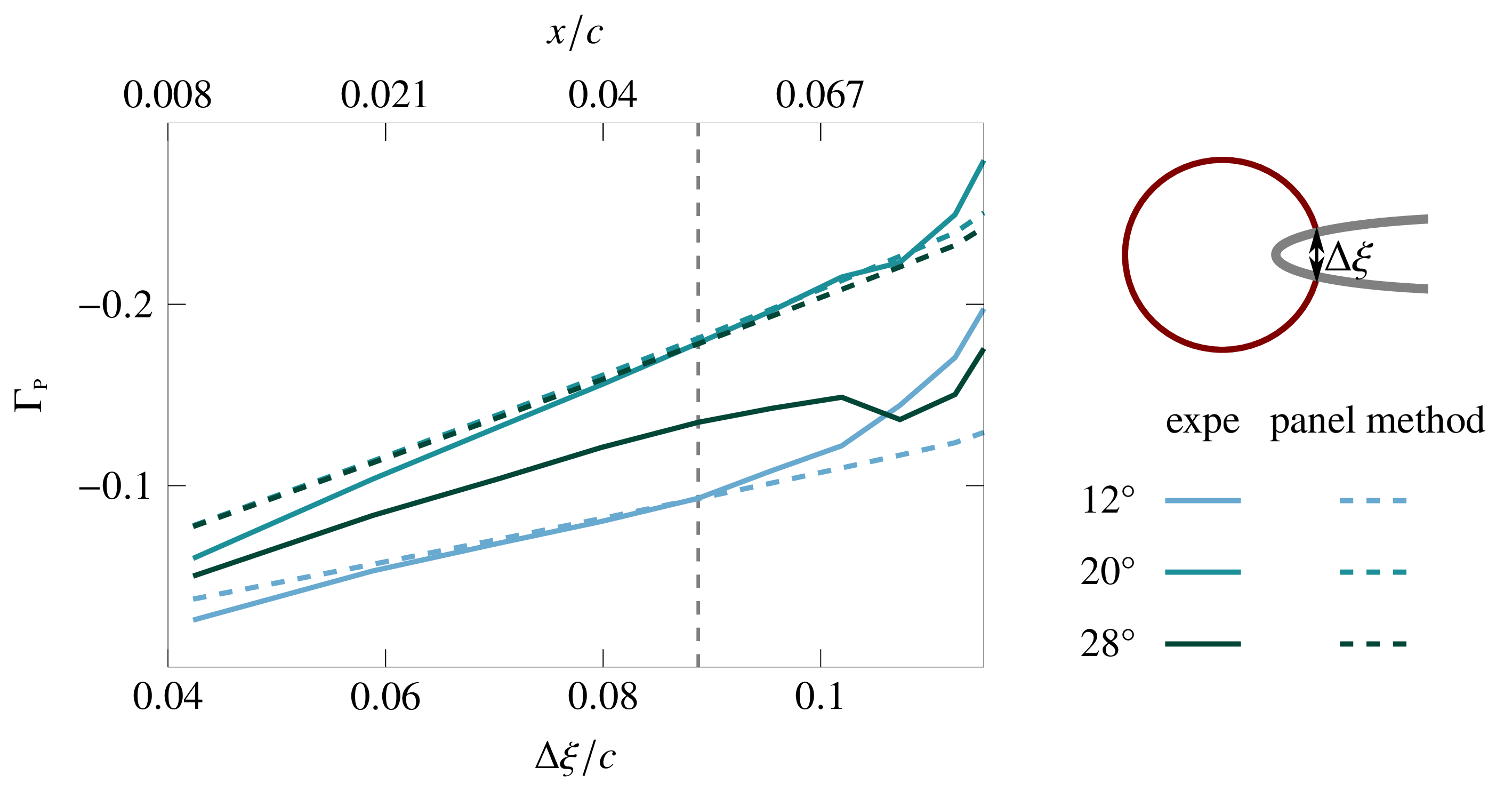}
\caption{Partial circulation $\kindex{\varGamma}{p}$ as a function of the distance between the endpoints of the integration contour for a pitching NACA0015 aerofoil.
Data is presented at three selected angles of attack: at the start of the pitching cycle: $\alpha=$\ang{12}, after a quarter period: $\alpha=$\ang{20}, and at mid-period: $\alpha=$\ang{28}.
The solid lines indicate data obtained from experimental flow field.
The dashed lines indicate values obtained using a panel method with a closing condition set at the stagnation point in the leading-edge region.
The axis on top indicates the chord-wise location of the endpoints of the integration contours.
}
\label{fig:PartialCircnaca0015}
\end{figure}

Following the analysis described in \cref{subsec:validationLESP} and summarised in \cref{fig:PartialCirc}, we verify first the dependence of the leading-edge suction parameter on the size of the integration contour.
\Cref{fig:PartialCircnaca0015} shows the variation of the partial circulation with increasing size of the integration contours as a function of the thickness of the aerofoil at the endpoints of the contours for three selected time instances during the pitching motion.
The selected snapshots are at the minimum angle of attack $\alpha = \ang{12}$, at the static stall angle $\alpha = \ang{20}$ during the pitch-up motion, and at the maximum angle of attack $\alpha = \ang{28}$.
At the two lower angles of attack, the flow is still attached and at the maximum angle of attack the flow is massively separated.
When the contour is too small ($x/c < 0.02$), the increase in the partial circulation with $\Delta\xt$ deviates slightly from the expected linear evolution predicted by the panel method for $\alpha = \ang{12}$ and $\alpha = \ang{20}$.
This is attributed to the limited resolution of the PIV data and shadow areas and noise near the surface of the aerofoil close to the very leading-edge.
For $0.02 < x/c < 0.05$, the partial circulation increases linearly for all angles of attack and unique values of the partial circulation can be extracted.
For larger contours, the partial circulation deviates again from the linear increase.
A panel method that uses the Kutta condition set at the leading-edge stagnation point correctly estimates the partial circulation at $\alpha = \ang{12}$, provides a fair estimation for $\alpha = \ang{20}$ when the flow starts to separate near the trailing edge, but overestimates the partial circulation when the flow is completely separated at $\alpha = \ang{28}$.
The partial circulation estimated by the panel method at $\alpha =\ang{28}$ equals the value estimated at $\alpha =\ang{20}$ because it takes into account the experimentally obtained location of the leading edge stagnation point which has moved towards the leading edge post dynamic stall.
For the panel method, this looks like a decrease in the effective angle of attack.
The conclusions are similar to those presented in \cref{subsec:validationLESP}.
To minimise the influence of the noise and shadows near the surface, the largest possible contour with the endpoints at \SI{5}{\percent} of the chord is chosen for the computation of the partial circulation.

\begin{figure}
\centering
\includegraphics{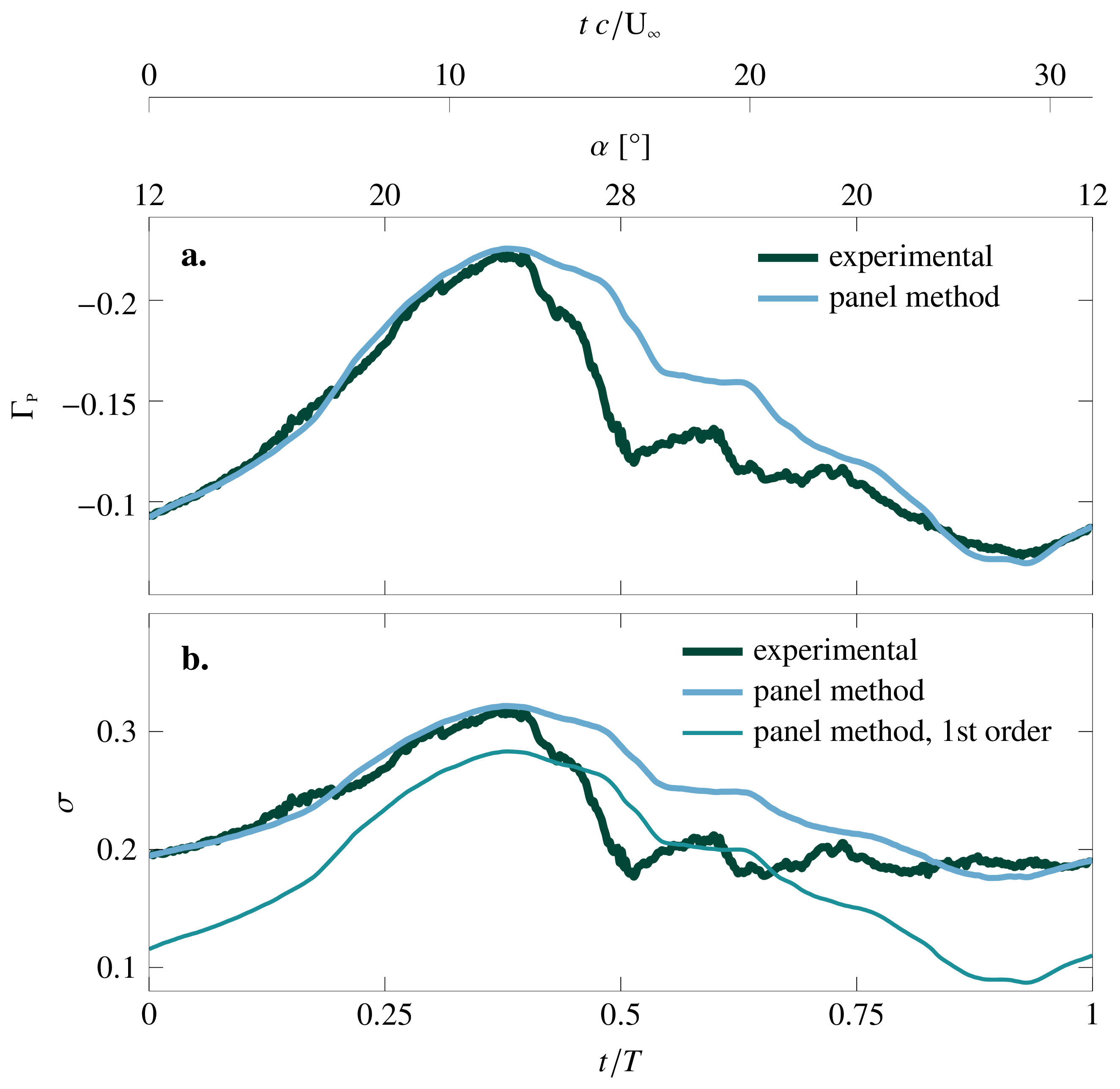}
\caption{Evolution of the partial circulations (a.) and the leading-edge suction parameters (b.) from flow field measurement and panel method.}
\label{fig:pitchingLESP}
\end{figure}

The evolution of the experimentally measured circulation at the leading edge is presented in \cref{fig:pitchingLESP}a.
The partial circulation at the leading edge using the experimental flow field can be compared with the panel method similarly to what we did previously in \cref{subsec:validationLESP}.
A standard panel method with a Kutta condition at the trailing edge would provide partial circulation values far away from experimental values in this case.
For example, we would obtain values of \num{-0.14} instead of \num{-0.09} at \ang{12} and \num{-0.34} instead of \num{-0.12} at \ang{28}.
Instead of imposing the Kutta condition at the trailing edge, we close the panel method conditions by imposing the leading-edge stagnation point at its measured location.
The resulting solution of the panel method then corresponds well to the experimentally measured values when the flow is attached.

The partial circulation calculated from the panel method with a Kutta condition at the stagnation point in the leading-edge region is shown by the light blue line in \cref{fig:pitchingLESP}.
It fits the experimental partial circulation until the maximum of the circulation is reached.
When the leading-edge vortex is being formed, and the experimental partial circulation quickly drops, the partial circulation predicted by the panel method no longer follows the experimental curve.
As long as the flow is attached at the leading edge, the circulation generated by the leading edge is well estimated by a potential flow if the stagnation point position is set.
Similar to the partial circulation, the leading-edge suction parameter $\sigma$, from the panel method coincides with the experimental one until the maximum is reached (\cref{fig:pitchingLESP}b).

\Cref{fig:pitchingLESP}b. also shows the leading-edge suction parameter using the simplified equation \cref{eq:lesp2}.
At $\alpha=\ang{12}$, the leading-edge suction parameter is almost half of the value determined using the entire \cref{eq:lesp1}.
For thick aerofoils, such as the NACA0015, the chord-wise position of the stagnation point is of the same order of magnitude as the leading-edge radius, and must be taken into account in the calculation of the leading-edge suction parameter.


When the flow starts to detach in the leading-edge region, the leading-edge suction parameter estimated by potential flow does not correspond to the measured value anymore.
The leading-edge suction parameter from potential flow mostly follows the slow dynamics of the position of the stagnation point, and does not reflect the important role of the leading-edge shear layer.
The shear layer height in the leading-edge region is correlated to the amount of the flow separation, which cannot be easily quantified from experimental flow field measurements, but can be quantified by the difference between the experimental and the computed leading-edge suction parameters.

\subsection{Estimation of the shear layer height in the leading-edge region}

\begin{figure}
\centering
\includegraphics{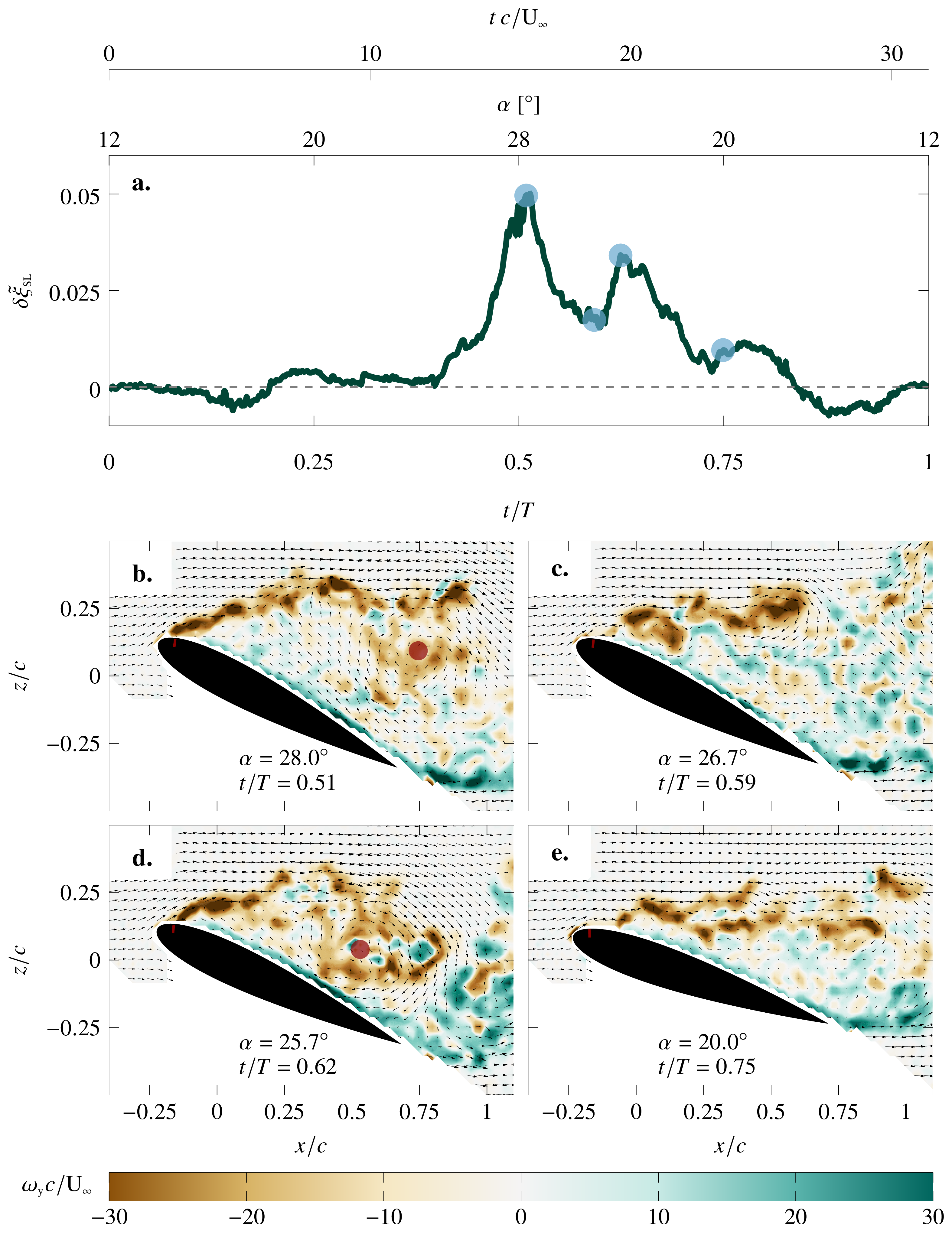}
\caption{a. Evolution of the estimated shear layer height using a comparison between the experimental and potential leading-edge suction parameters. The four markers correspond to the four snapshots presented below at maximum of shear layer height (b.), at a local minimum. (c.), at the second local maximum. (d.) and during the decrease of the shear layer height before reaching zero (e.). The red mark at the leading edge of the aerofoil indicates the \SI{5}{\percent} chord length, and the red dot represents the centre of the leading-edge vortex.}
\label{fig:pitchingFlowLE_SL}
\end{figure}

The results presented in \cref{subsec:SLflowpert} indicated that the difference between the measured values and the potential flow based predictions of the leading-edge suction parameters comes predominantly from the shear layer height.
We will use this results here to deduce the shear layer height at the extremity of the partial contour (here at \SI{5}{\percent} of the chord) using \cref{eq:lespdx}.
The estimated shear layer height is presented in \cref{fig:pitchingFlowLE_SL}a.
The shear layer height stays around zero until $t/T=0.4, \alpha=\ang{26}$.
Hereafter, the shear layer height quickly increases from zero to \num{0.05}$x/c$ in three convective times to reach a maximum at half of the pitching period.
After a sharp decrease of the shear layer height, a second maximum is observed around $t/T = 0.62$.
After this second local maximum, the shear layer decreases and reaches zero again at the end of the pitching period, indicating flow reattachment.

To validate these results, we present flow field snapshots for four selected time instants indicated by the markers in \cref{fig:pitchingFlowLE_SL}a.
The red marker, at the leading-edge of the aerofoil, indicates the \SI{5}{\percent} chord position on the suction side, where the shear layer height is estimated.
At the time instant when the shear layer height is maximum, a coherent leading-edge vortex can be observed with a vortex centre directly above the trailing edge (see red dot in \cref{fig:pitchingFlowLE_SL}b).
Shortly after, at the beginning of the pitch-down movement, the flow is mostly detached over the aerofoil but locally at the leading edge, the flow is reattached (\cref{fig:pitchingFlowLE_SL}c.), corresponding to a local minimum of the estimated shear layer height.
In \cref{fig:pitchingFlowLE_SL}d., a second smaller leading-edge vortex is seen above the half aft part of the aerofoil at three quarter of the chord, similarly to the main leading-edge vortex in \cref{fig:pitchingFlowLE_SL}b.
The induced velocity of this secondary leading-edge vortex pushes the shear layer height up and explains the local maximum in the evolution of the shear layer height.
At $\alpha=\ang{20}$ on the downstroke, the flow starts to reattach at the leading edge (\cref{fig:pitchingFlowLE_SL}e.).
Remnant of detached flow are present over most of the aerofoil, but the flow around the leading edge has started to reattach, which corresponds to a decrease of the shear layer height in \cref{fig:pitchingFlowLE_SL}a.

The shear layer height estimation from the leading-edge suction parameter corresponds to what is observed on the flow field at the leading edge.
The shear layer height is able to detect when the flow is attached at the leading edge but detached over the rest of the aerofoil.
A maximum of shear layer height at the leading edge corresponds to a detachment of the leading-edge vortex from the aerofoil and would be a direct indicator of dynamic stall onset following the definition of onset proposed in \cite{mulleners_onset_2012}.
The shear layer height estimation allows a direct surface based observation of the state and the degree of flow separation on an unsteady moving aerofoil or an aerofoil in an unsteady flow environment.

\section{Conclusions}																	

A leading-edge suction parameter has been mathematically defined using a potential flow model passing around an edge.
The leading-edge suction parameter is defined based on the partial circulation around the leading edge, the leading-edge radius, and the position of the stagnation point.
Here, we have extracted the leading-edge suction parameter from computationally and experimentally obtained flow field data.
To compute the partial circulation, we selected circular arc-shape integration contours that intersect perpendicularly with the aerofoil contour and we normalise the values of the partial circulation by the normal distance between the two symmetric endpoints of the contour to obtain a value for the leading edge suction parameter that is independent of the size of the integration contour.

For thin aerofoils, such as a NACA0009, we demonstrate that we indeed obtain a unique value of the leading-edge suction parameter if the contour encloses up to \SI{10}{\percent} of the chord.
The small leading-edge radius of these aerofoils makes it possible to simplify \cref{eq:lesp1} to \cref{eq:lesp2}, where information on the stagnation point location is not required to determine the leading-edge suction parameter.
The definition of the leading-edge suction parameter remains valid when the flow detaches from the aerofoil at the trailing edge because its definition does not assume a Kutta condition at the trailing edge, unlike classical thin-aerofoil theory.

For a thicker aerofoil, such as a NACA0015, the chord-wise position of the stagnation point is of the same order of magnitude as the leading-edge radius and must be taken into account to calculate the leading-edge suction parameter.
A panel method, with no boundary layer model, can still predict the measured leading-edge suction parameter, if the zero vorticity point in the panel method is set at the leading-edge stagnation point location instead of at the trailing edge.
The resulting leading-edge suction parameter predicts well the measured value even for highly unsteady flow conditions that occur during dynamic stall on a pitching aerofoil.

When the flow over the aerofoil separates near the leading-edge, it affects the leading-edge suction parameter.
The evolution of the separating shear layer can be considered as a local increase of the aerofoil thickness near the leading edge.
By adding the instantaneous value of the shear layer height to the thickness of the aerofoil, we can adjust the potential flow model to obtain accurate estimates of the experimentally computed leading-edge suction parameter.
The tight relationship between the leading-edge suction parameter, the partial circulation, and the shear layer height makes it possible to describe the suction generated at the leading edge and its evolution using a potential flow model, even when the flow over the aerofoil is separated.

We show that it is also possible to estimate the shear layer height using the partial circulation determined experimentally and the leading-edge suction parameter estimated with potential flow.
This means that we can experimentally quantify the degree of flow separation at the very leading edge of an aerofoil based on local flow measurements and identify the detachment of the leading-edge vortex from the aerofoil. 
Therefore, this approach has the potential to inspire data-driven model development, flow sensing, and control methods for applications related to gust alleviation and stall mitigation.

\backsection[Acknowledgements]{
We thank Dr. Guosheng He, now associate professor at the Beijing Institute of Technology, who carried out the experiments of the pitching aerofoil, with Lars Siegel and Dr. Arne Henning from the German aerospace center (DLR) in Göttingen.
We also thank Guillaume de Guyon from UNFoLD EPFL for the development of the panel method and the discussion which ended with the idea of the closing condition using the stagnation point at the leading edge.
We would like to thank Dr. Katherine Asztalos (now at the Argonne National Laboratory) and Professor Scott Dawson from the Illinois Institute of Technology for sharing the DNS and the panel method data for the NACA0009 aerofoil.
}
\backsection[Funding]{
The work presented is supported by the SNSF Lead Agency programme (K.M., grant number 200021E-169841); the SNSF Assistant Professor energy (K.M., grant number PYAPP2\_173652); and the U.S. Air Force Office of Scientific Research grant with program officer Gregg Abate (J.E. \& D.W., grant number FA9550-18-1-0440).
}

\backsection[Declaration of interests]{The authors report no conflict of interest.}

\backsection[Author ORCID]{J. Deparday, https://orcid.org/0000-0001-8040-546X;
X. He, https://orcid.org/0000-0003-3824-9744;
J. Eldredge, https://orcid.org/0000-0002-2672-706X;
K. Mulleners, https://orcid.org/0000-0003-4691-8231
}

\appendix

\section{Description of the parabola approximation}\label{app:Parabola}
\noindent
We approximate the leading edge of the aerofoil by a parabola with a radius of curvature \rle, described mathematically by
$\yt = \xt^{2}/(2\rle)$.
We will use the complex form of coordinates $\zt = \xt + \upi \yt$.
The $\yt$ axis represents the line of symmetry of the parabola, and $\zt = 0$ is always the nose of the parabola (\cref{fig:contour}).
By defining the conformal mapping
\begin{equation}
    \zt(\kappa)=\kappa+\frac{\upi}{2} \frac{\kappa^{2}}{\rle},
\label{eq:conform}
\end{equation}
the parabola shape is mapped from the real $\chi$ axis. Denoting $\kappa = \chi + \upi \psi$, then this axis is $\psi = 0$.

The Jacobian of this mapping is
\begin{equation}
\d \zt_{\kappa} = 1 + \upi \kappa/\rho,
\end{equation}
and the inverse of the mapping is
\begin{equation}
\kappa(\zt) = \upi \rle \left[ 1 - \left( 1 + \upi 2 \zt/\rle\right)^{1/2}\right].
\label{eq:inversemapping}
\end{equation}

This mapping is singular at a point inside the envelope of the parabola at $\kappa= \upi \rle$, where $\zt = \upi \rle/ 2$.
We designate the branch cut of the inverse to follow the $\yt$ axis above this point.
The entire mapping can be easily rotated by angle $\alpha$ in the physical $z$ plane by defining $z = \zt \exp(\upi \alpha)$.

To model a uniform flow past this parabola, we write the following complex potential in the $\kappa$ plane:
\begin{equation}
\hat{F}(\kappa) = A \left(\kappa - a\sqrt{2\rle}\right)^{2}.
\label{eq:potential}
\end{equation}
This potential represents a basic stagnation flow, with a stagnation point on the $\chi$ axis at $\kappa_{s}=a \sqrt{2 \rle}$ and stagnation streamlines parallel to the $\chi$ and $\psi$ axes.
The factor $A$ represents the strength of this stagnation flow.

This flow is mapped to the $\zt$ plane via the inverse conformal mapping~\cref{eq:inversemapping}:
\begin{equation}
    F(\zt) = \hat{F}(\kappa(\zt))
\end{equation}
The complex conjugate velocity in the $\zt$ plane follows from the derivative of this potential,
\begin{equation}
w(\zt) = u - \upi v = \d F_{\zt} = \d \hat{F}_{\kappa} \left(\d \zt_{\kappa}\right)^{-1} = 2A \left(\frac{\kappa(\zt) - a\sqrt{2\rle}}{1 + \upi \kappa(\zt)/\rle}\right)
\end{equation}
We have written this complex velocity in terms of $\kappa$ for tidiness; it is easily written in terms of $\zt$ by substituting the inverse transform~\cref{eq:inversemapping}.
It is easy to verify that the stagnation point $\kappa_{s}=a \sqrt{2 \rle}$ is mapped to $\zt_{s}=a \sqrt{2 \rle}+\mathrm{i} a^{2}$ on the parabola.

This flow is well behaved everywhere.
In fact, as $|\kappa| \rightarrow \infty$ (or, equivalently, as $|\zt| \rightarrow \infty$), the velocity approaches $-\upi 2 A \rle$, which represents a uniform flow of strength $U = 2 A \rle$ in the $\yt$ direction, parallel to the axis of symmetry. Let us write the complex velocity again, now in terms of $U$:
\begin{equation}
w(\zt) = U \left(\frac{\kappa(\zt)/\rle - a\sqrt{2/\rle}}{1 + \upi \kappa(\zt)/\rle}\right) = -\upi U \left[ 1 - \frac{1 + \upi a \sqrt{2/\rle}}{\left( 1 + \upi 2 \zt/\rle\right)^{1/2}}\right].
\label{eq:vel}
\end{equation}
The pressure distribution about the parabola follows from the Bernoulli equation,
\begin{equation}
    p(\zt) - p_{\infty} = -\frac{1}{2} \rle |w(\zt)|^{2},
\end{equation}
where $|\cdot|^{2}=w w^{*}$ and $(\cdot)^{*}$ denotes the complex conjugate.
We assume here that the flow is steady for simplicity, but the main contributions of this approach are not dependent on this assumption.

\bibliographystyle{jfm}
\bibliography{2021_lesp}

\end{document}

%% file: figurematter/plots/SketchThicknessSL.pdf_tex

\begingroup%
  \makeatletter%
  \providecommand\color[2][]{%
    \errmessage{(Inkscape) Color is used for the text in Inkscape, but the package 'color.sty' is not loaded}%
    \renewcommand\color[2][]{}%
  }%
  \providecommand\transparent[1]{%
    \errmessage{(Inkscape) Transparency is used (non-zero) for the text in Inkscape, but the package 'transparent.sty' is not loaded}%
    \renewcommand\transparent[1]{}%
  }%
  \providecommand\rotatebox[2]{#2}%
  \newcommand*\fsize{\dimexpr\f@size pt\relax}%
  \newcommand*\lineheight[1]{\fontsize{\fsize}{#1\fsize}\selectfont}%
  \ifx\svgwidth\undefined%
    \setlength{\unitlength}{66.68773621bp}%
    \ifx\svgscale\undefined%
      \relax%
    \else%
      \setlength{\unitlength}{\unitlength * \real{\svgscale}}%
    \fi%
  \else%
    \setlength{\unitlength}{\svgwidth}%
  \fi%
  \global\let\svgwidth\undefined%
  \global\let\svgscale\undefined%
  \makeatother%
  \begin{picture}(1,0.63183563)%
    \lineheight{1}%
    \setlength\tabcolsep{0pt}%
    \put(0,0){\includegraphics[width=\unitlength]{./figurematter/plots/SketchThicknessSL.pdf}}%
    \put(0.79,0.32){\color[rgb]{0,0,0}\makebox(0,0)[lt]{\lineheight{1.25}\smash{\begin{tabular}[t]{l}$\Delta \xi$\end{tabular}}}}%
    \put(0.79,0.39){\color[rgb]{0,0,0}\makebox(0,0)[lt]{\lineheight{1.25}\smash{\begin{tabular}[t]{l}$\delta\kindex{\xi}{SL}$\end{tabular}}}}%
  \end{picture}%
\endgroup%